\documentclass{aa}  
\usepackage{graphicx}
\usepackage{txfonts}
\usepackage{changes}
\usepackage{natbib}
\usepackage{pdflscape}
\usepackage{xcolor}	
\usepackage{longtable}
\usepackage{subcaption}
\usepackage{tabularx}

\usepackage{hyperref}
\hypersetup{colorlinks=true,linkcolor=blue,citecolor=blue,filecolor=blue,urlcolor=blue}
\newcommand{\msun}{\ensuremath{\, {M}_\odot}}
\newcommand{\Msun}{\ensuremath{\, {M}_\odot}}

\begin{document} 

   \title{GN-z11: witnessing the formation of second generation stars and an accreting massive black hole in a massive star cluster}
   \authorrunning{F. D'Antona et al.}
   \titlerunning{The status of GN-z11 and the role of AGB stars }
%  \subtitle{I. Gas }

   \author{F. D'Antona\inst{1}, E. Vesperini\inst{2}, F. Calura\inst{3}, P. Ventura\inst{1}, A. D'Ercole\inst{3}, V. Caloi\inst{4}, A. F. Marino\inst{5,6}, A. P. Milone\inst{5,7}, \\ F. Dell'Agli\inst{1} and M. Tailo\inst{8}
          }

   \institute{
              Istituto Nazionale di Astrofisica, Osservatorio Astronomico di Roma, Via Frascati 33, 00077 Monte Porzio Catone, Italy 
              \and Department of Astronomy, Indiana University, Swain West, 727 E. 3rd Street, IN 47405 Bloomington (USA) 
              \and INAF -- OAS, Osservatorio di Astrofisica e Scienza dello Spazio di Bologna, via Gobetti 93/3, I-40129 Bologna, Italy
              \and Istituto Nazionale di Astrofisica -- Istituto di Astrofisica e Planetologia Spaziali, Via Fosso del Cavaliere 100, I-00133, Roma, Italy 
              \and Istituto Nazionale di Astrofisica -- Osservatorio Astronomico di Padova, Vicolo dell'Osservatorio 5, Padova, I-35122             
             \and Istituto Nazionale di Astrofisica, Osservatorio Astrofisico di Arcetri, Largo Enrico Fermi, 5, Firenze, IT 50125        
              \and Dipartimento di Fisica e Astronomia ``Galileo Galilei'', Univ. di Padova, Vicolo dell'Osservatorio 3, Padova, I-35122
             \and Dipartimento di Fisica e Astronomia Augusto Righi, Universit\`a degli Studi di Bologna, Via Gobetti 93/2, I-40129, Bologna, Italy     
             }

 %  \date{Received September 15, 1996; accepted March 16, 1997}

% \abstract{}{}{}{}{} 
% 5 {} token are mandatory
 
  \abstract
  % context heading (optional)
  % {} leave it empty if necessary  
 %  {Star formation and growth of a central black hole in a possible nuclear star cluster at high redshift.}
  % aims heading (mandatory)
  % { We explore the possibility that we are observing the formation of the second generation, according to the AGB model, in a massive Globular Cluster, or in a Nuclear Star Cluster in the young proto-galaxy GN-z11, observed at a redshift z=10.6.}
  % methods heading (mandatory)
 %  {We show that a second generation forming in the accreting gas polluted by the ejecta of massive Asymptotic Giant Branch (AGB) stars and mixed with gas having a standard composition accounts for the unusually large N/O in GN-z11 spectrum. The timing of the evolution of massive (4--7.5\Msun) AGBs provides a favourable environment for the growth of a central stellar mass black hole to the present day observed Active Galactic Nucleus stage.}
  % results heading (mandatory)
  % {The progenitor system was born at an age of the Universe of  $\sim$260--380\,Myr, well within the pre--reionization epoch.  }
  % conclusions heading (optional), leave it empty if necessary 
  % {}
{We explore the possibility that the  N-rich young proto-galaxy GN-z11 recently observed at z=10.6 by the James Webb Space Telescope  is the result of the formation of second generation stars from pristine gas and Asymptotic Giant Branch (AGB) ejecta in a massive globular cluster or nuclear star cluster. We show that a second generation forming out of gas polluted by the ejecta of massive AGB stars and mixed with gas having a standard composition accounts for the unusually large N/O in the GN-z11 spectrum. The timing of the evolution of massive (4--7.5\Msun) AGBs also provides a favourable environment for the growth of a central stellar mass black hole to the Active Galactic Nucleus stage observed  in GN-z11. According to our model the progenitor system was born at an age of the Universe of $\simeq 260 - 380$\,Myr, well within the pre-reionization epoch.}

   \keywords{Stars: evolution; Stars: AGB and post-AGB;  Stars: black holes; Stars: formation; globular clusters: general; galaxies: abundances; galaxies: high-redshift; quasars: supermassive black holes
              }

   \maketitle
%
%-------------------------------------------------------------------
\begin{figure}
	% To include a figure from a file named example.*
	% Allowable file formats are eps or ps if compiling using latex
	% or pdf, png, jpg if compiling using pdflatex
%	\includegraphics[width=15cm]{./figurefinal/figure1letter.pdf}
\vskip -40pt  
\begin{minipage}{0.48\textwidth}
\resizebox{1.\hsize}{!}{\includegraphics{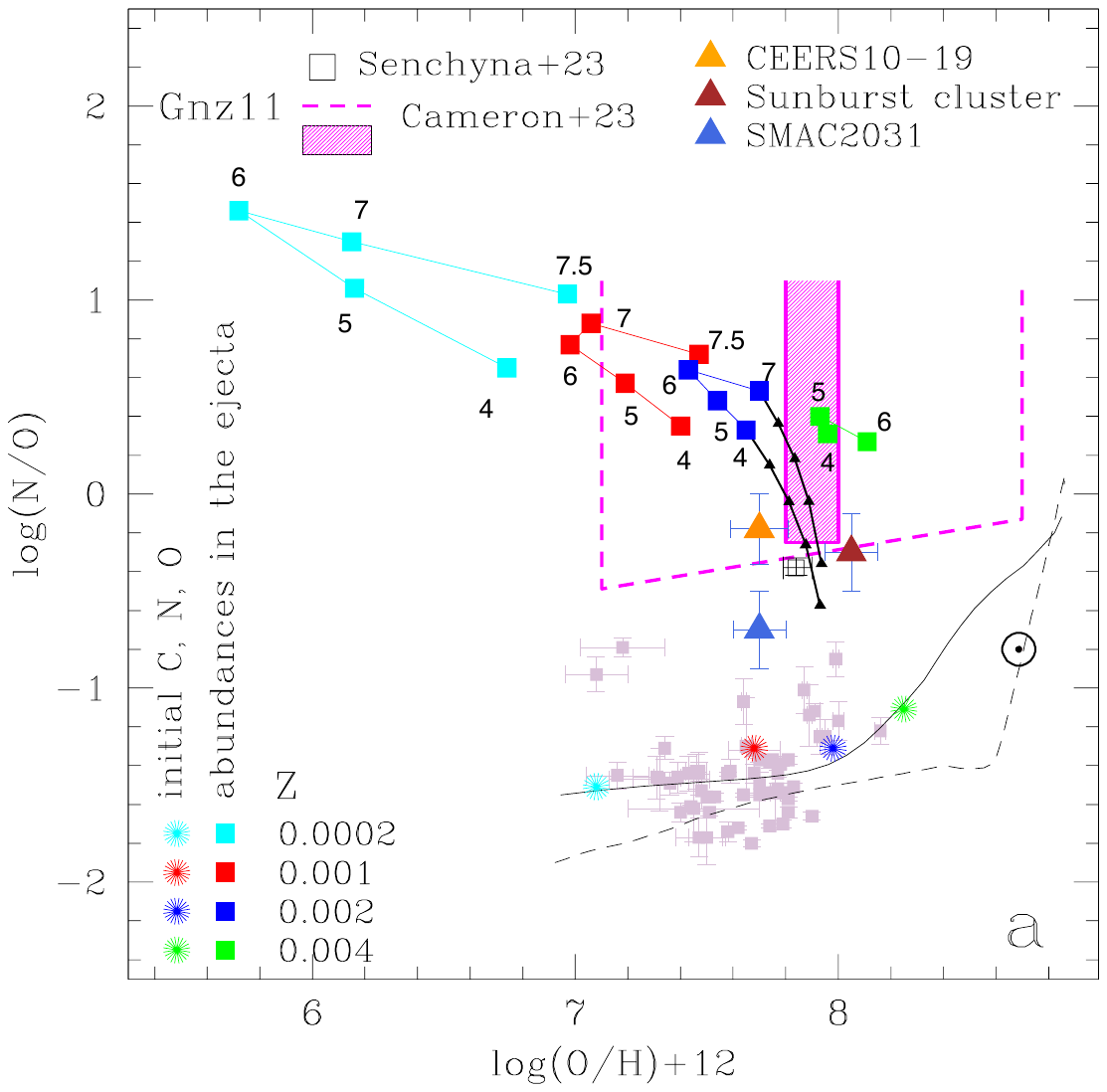}}	 
\vskip -135pt  
\resizebox{1.\hsize}{!}{ \includegraphics{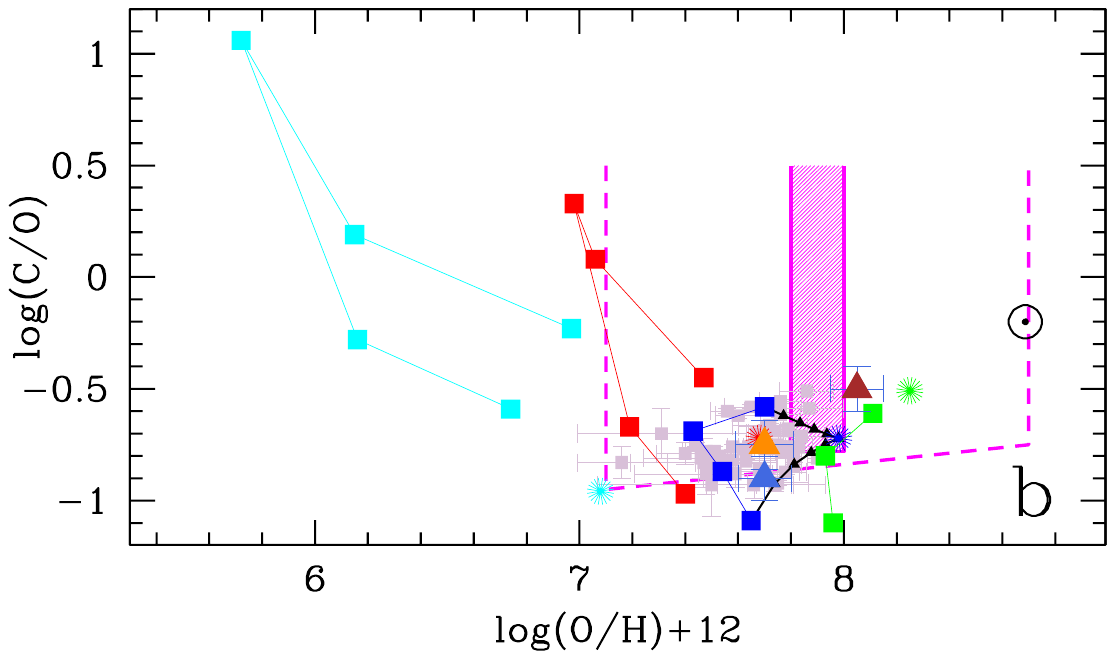}}	 
\end{minipage}
\vskip -168pt  
\caption{ Abundances in GN-z11 and other objects, compared with models. Panel a: diagram log(N/O) versus log(O/H)+12: ``fiducial" (pink filled) and ``conservative" (pink dashed border) error boxes for  GN-z11 \citep{cameron2023}.  Black open square with error bars: abundances from \citet{senchyna2023}. Colored squares:  average abundances ratios in the ejecta of intermediate mass AGB stars \citep{ventura2013}. Metallicities in mass fraction Z and  masses (from 7.5 to 4\msun)  are labelled in the figure. Asterisks: initial O/H and N/O  abundances of the models (see Table 1 and \citet{ventura2013}). Solar symbol is placed at the values corresponding to solar abundances.
%, based both on the metallicity and on the chosen [$\alpha$/Fe] enhancement  (see Table 1 and \citet{ventura2013} for details). 
Black lines: examples of dilution of the ejecta (of the 4 and 7\Msun at Z=0.002) with gas having the initial chemical composition; triangles are marking 20, 40, 60 and 80\% of diluting pristine gas.
Thistle squares with error bars: abundances observed in low metallicity galaxies at z=0.3-0.4 \citep{izotov2023}  and in the  local compact galaxies  from the SLOAN \citep{izotov2012}. At the bottom: upper (full)  and lower (dashed) envelope  of the chemical evolution models by \citet{vincenzo2016}.  Other star forming galaxies at high redshift having high log(N/O) are shown as colored triangles and are identified in the labels.
Panel b: diagram  log(C/O) versus log(O/H)+12. Symbols and lines as in panel a. 
}  
\label{fig:1}       
\end{figure}

\section{Introduction}
The recent observations of GN-z11 are opening an unexpected window on our knowledge of the first phases of evolution of the Universe. The precise redshift measurement of z=10.603 from JWST/NIRSpec \citep{bunker2023} dates it at 430\,Myr in the $\Lambda$CDM model, based on the \cite{planckcoll2020} determination of cosmological parameters. In the JADES NIRCam imaging \citep{tacchella2023}, this object shows an intrinsic half-light radius of only 0.0016$\pm$0.005$"$ (64$\pm$20\,pc), suggesting that it may host a possible Globular Cluster (GC) in formation \citep{senchyna2023, belokurov2023}. Such a low age poses a puzzle, as  the spectrum reveals a very high nitrogen abundance, not seen at similar metallicity in the nearby Universe \citep[e.g.][]{izotov2012, izotov2023}, nor predicted by standard chemical evolution models \citep[e.g.][]{vincenzo2016}, while a high [N/O] is consistent with the abundances in the ``anomalous'' Globular Cluster stars, suggesting that we may be witnessing the formation of their ``second generation" \citep{charbonnel2023, renzini2023, marques2023}.

\citet{cameron2023} quantify the relative abundances of C, N and O from the spectrum of GN-z11, and provide error boxes for both ``fiducial" abundances and for more ``conservative" assumptions. In both cases, the most relevant result is the lower limit to the Nitrogen to Oxygen ratio, log[N/O]$>$--0.25  \citep[a value also compatible with the determination  from detailed nebular models by][]{senchyna2023}, more than one order of magnitude higher than the values found in galaxies of a similar low metallicity (log(O/H)+12$\lesssim$8.0) at low redshift and also larger than the solar value (log(N/O)=$-0.8$). The Nitrogen and Oxygen observed in star forming galaxies in the local Universe is well reproduced by chemical evolution models \citep{vincenzo2016}, while very ad hoc assumptions should be made to account for such an anomalous and fast Nitrogen increase observed in the spectrum of the young and compact GN-z11 \citep{bekki2023, nagele-umeda2023, kobayashi2023}. Nevertheless, these same abundances are common among a great fraction of the stars of Globular Clusters (GC), where indeed peculiar abundance patterns apparently emerged very early in the life of the Universe, in which the gas had been processed by proton capture reactions at high temperature (T$>$40\,MK), leaving the signatures of full CNO processing and of the Ne--Na and Mg--Al chains \citep[e.g.][]{gratton2019}. The main process invoked in the  formation of GC ``multiple populations" is that this hot H--burning occurs inside the stars formed at first in the cluster, and that the products of these nuclear reactions, collected and mixed with non--processed gas, formed a ``second generation" which typically represents the most abundant component  \citep[from $\sim$45\% to 90\% of stars, see e.g.][]{milonemarino2022} of today's GCs. Thus the birth of the second generation is a major event in the star formation history of GCs. \\
Similarly high N/O abundances are found (see Fig.\,\ref{fig:1}) in a few other star forming objects at lower resdshift, CEERS-1019 at z$\sim$8.7 (Isobe et al. 2023), \citep{isobe2023}, 
the Super star Cluster in the Sunburst galaxy \citep{pascale2023} (at z=2.67), a galaxy at z=3.5 lensed by the cluster SMACS2031, the Lynx arc at z=3.37 \citep{marques2023}. For some of these objects too it has been suggested that they could be hosting the formation of second generation in GCs \citep{isobe2023, marques2023}.
Although an high N/O ratio is an important feature characterizing  the abundances in the second generation stars in GCs, it can result from a variety of mass loss events where CNO cycled material is expelled, such as stages of massive Wolf--Rayet \citep{limongi2018, kobayashi2023}, common envelope or other forms of strong mass loss in the non conservative evolution evolution of interacting binaries \citep{demink2009}, mass loss from rapidly rotating massive stars \citep{decressin2007}, and any of these explanations may be apt to describe the present status of the gas in these galaxies, especially in the very young Sunburst galaxy \citep{pascale2023}. In all these cases, we should be observing a very short phase in the life of the forming galaxy. It is important to emphasize that a model for the formation of second generation stars in GCs must explain several stringent and more complex chemical constraints such as Na-O and Mg-Al anticorrelations, the Lithium abundances, the high helium populations, the maximum value of the helium, the discreteness in the distribution of chemical anomalies, in addition to a variety of  other minor but important properties (see \cite{gratton2019} for a comprehensive discussion of all these issues).  These constraints are so strong that most of models fail to comply with them, and in fact the supermassive stars model (SMS) had been originally developed mainly to deal with the depletion of $^{24}$Mg implied by the Mg--Al anticorrelation \citep{denissenkovherwig2003,denissenkov2014}. Although with some limitations \citep{renzini2015}, the Asymptotic Giant Branch (AGB) model \citep{dercole2008, dercole2016, calura2019} in particular deals well with most of the abundance trends, including \citep{ventura2012}  the extreme Mg--K anticorrelation found in a few clusters \citep{cohenkirby2012, mucciarelli2015, carretta2022} and with the variety and discreteness of multiple populations \citep{dantona2016}.

The main aim of this letter is to show that the composition of AGB ejecta, diluted with gas having the composition of the first generation, is compatible with the high N/O found in the spectrum of GN-z11. 
Simultaneously, it aims to highlight that the long phase of massive AGB evolution is consistent with the time needed to build up the central accreting black (BH) hole of mass $\gtrsim 10^6$\Msun\ apparently hosted in the system \citep{maio2023bh106}. We remark that the model proposed here represents just a plausible scenario to explain several features of this peculiar object simultaneously, but it is not necessarily the unique solution. \\
The outline of the paper is the following: we start with examining the problem of the timescales in Sect.\,\ref{sec2}, we then proceed to compare the abundances in Sect.\,\ref{sec3} and we conclude with a discussion of  unsolved issues along with possible promising approaches in Sect.\,\ref{sec4}. In Sect.\,\ref{sec5} we compare the timescales for the possible formation route to build the massive BH with the timescale of the AGB evolution, and summarize the results in  Sect.\,\ref{sec6}.

\section{Timings involved: the central Black Hole}
\label{sec2}
If we were witnessing the formation of a second generation, the main difference between the AGB model and all the others is the timescale on which it operates. The event  ranges from less than a million year if the source of Nitrogen rich gas are SMS \citep{charbonnel2023}, up to as long as 10\,Myr if the polluters are massive binary interactive stars possibly operating  in a `quiet' epoch, when masses $\gtrsim 25$\Msun\ evolve  directly to black holes, as recently suggested by  \cite{renzini2023}. Instead, the massive AGB evolution timescale is of the order of several tens of million years, a good reason to raise criticism to their possible role.  \citet{senchyna2023} note that the 40 to more than 100\,Myr  of the AGB evolution are difficult to accomodate into the  $\sim 20$\,Myr age of the burst we are looking at. According to the analysis of the JADES NIRcam imaging \citep{tacchella2023}, half of the star formation in the central regions took place in the last few million years, while an extended star formation epoch at a much smaller rate may have occurred both in the central point and in the extended source. Also \cite{cameron2023} discarded the model, by noting that a fine tuning is required between the Nitrogen production and the lack of Oxygen increase associated with the SN\,II ejecta of  the intense star formation at 20\msun/yr.  Basically, such a star formation rate (SFR) leaves scarce space for the long operation of the AGB model, as a very massive second generation would be formed in much less than a million year. 
\\
On the other hand, the models on which these results are based did not include the possibility that the central regions host an Active Galactic Nucleus (AGN). 
As pointed out by \citet{maio2023bh106}, the observation in the spectrum of high ionization semi-forbidden nebular lines typical of the Broad Line Region of AGNs, and of blue shifted lines testifying the presence of fast outflows, provides a strong indication of the presence of a central accreting black hole (BH). The quantification of the intense star formation should take this into account.
Recently \citet{dsilva2023} revised the star formation history of galaxies  including in the fit an AGN component. They find  star formation histories a factor $\sim$9 lower than the fit not including the AGN for galaxies in the range $z >$9, and a reduction even up to 4\,dex  for  individual cases. Thus, the presence of an AGN component  in GN-z11 allows us to reconsider  both the determination of a total mass of $\sim 10^9$\Msun, and of the SFR of $\sim 20$\Msun/yr, and look under a different light to the possible timescale on which the N-rich sources act.

We take the point of view that the PS component is due to an AGN, while the extended component would be the host galaxy \citep{tacchella2023}.  Scaling of the AGN properties to the case of GN-z11 provides a mass of $\sim 10^6$\Msun\ for the central BH  \citep{maio2023bh106}, so that GN-z11 also poses an important age constraint on the formation of massive BHs \citep[see][for a general reference]{greene2020}. Such a massive BH could be the result of direct collapse of a gas cloud into BH \citep[see, e.g.][and references therein]{latif2016}. 
Otherwhise it must result from mass accretion on typical (20--40\Msun)\footnote{We adopt here the BH masses from the models by \citet{limongi2018} for [Fe/H]$= -1$, in the mass range 25--80\Msun.} stellar remnant BHs, or more massive seed BHs runaway stellar mergers \citep[see e.g][]{portegieszwart2002}.
A favourable environment for the latter scenario is met if we are indeed in the central region of a massive first generation GC. Many BHs would be the remnants of the evolution of the masses M$>$25\Msun, and  repeated stellar BHs early mergers may  increase the coalesced BH mass to $\sim$100\msun, or even to $\sim10^3$\Msun\ for the largest initial masses of the cluster \citep{rodriguez2019,antonini2019,kritos2023}. A BH mass of $\sim$10$^6$\Msun\ would eventually be reached if there is time for a ``long" phase of accretion. \citet{maio2023bh106} estimate 30--150\,Myr for this phase, starting from a mass of 100\Msun, at the Super--Eddington mass accretion rates $5\pm$2 times the Eddington rate implied by the AGN luminosity of 10$^{45}$erg/s.  Note that the presence of an accreting intermediate mass BH, atypical for standard GCs, makes GN-z11 more similar to the Nuclear Star Clusters (NSC) at the center of galaxies \citep{ferrarese2006, neumayer2020AAR}. 
If we are at the centre of a massive cluster, the time lapse for the accretion phase may be met during the long period of quiet following the Supernovae type\,II (SN\,II) epoch, during which the massive AGB winds accumulate in a central cooling flow and form the second generation stars, as envisioned by the AGB model \citep{dercole2008, dercole2016}, now supported by 3D hydrodynamical simulations \citep{calura2019, yaghoobi2022}. The growth of a central BH to masses up to $\sim 10^4$\Msun\ had been modelled in this context by \citet{vesperini2010bh}, considering only sub--Eddington or Eddington accretion rates on a relative low--mass seed BH ($\sim$100\Msun) possibly produced by early stellar mergers. If super--Eddington rates are allowed, the mass at z=10.6 may indeed become as large as $\sim$10$^6$\Msun, as required. Note anyway that the cooling flow of the AGB model does not require the presence of a central BH, as it is due to the gravitational well of the first generation cluster.\\
The AGB model for the formation of the second generation in GCs provides then a favourable set for the build up of the central BH, unlike other possible models, which must rely on the hypothesis of a BH massive from its birth.

\section{Comparison of abundances}
\label{sec3}
It is necessary to examine whether the chemistry of the AGB ejecta can reproduce the observed N/O and C/O abundances of the emission spectrum. In Figure \ref{fig:1} we plot the error boxes \citep{cameron2023} of GN-z11 in the log(N/O) and log (C/O) versus 12+log(O/H) planes. The results for N/O and O/H of the analysis made by examining  detailed nebular models \citep{senchyna2023} is reported as an open black square. 
The coloured squares in the plots represent the average abundances of the ejecta from the AGB models of different mass and metallicity, as listed in the figure \citep{ventura2013}. The abundances in the starting main sequence models are represented by the asterisks, and we see that they are typical abundances consistent with the values found for the compact local galaxies \citep{izotov2012}.  The black lines drawn between the AGB ejecta and the initial composition represent examples of the possible dilution of the AGB gas and gas with the primordial composition of the interstellar medium. Considering different dilution degrees, we can fit the abundances within the fiducial and/or conservative error box, after excluding the lowest metallicity models (Z=0.0002, or initial log(O/H)+12=7.68) clearly out of the errors. The highest metallicity (Z=0.004 or initial log(O/H)+12=8.04) is well within the fiducial error box with the ejecta of the 4 and 5\Msun. The two intermediate metallicity evolutions (Z=0.001 and Z=0.002) are both compatible with the composition of gas in GN-z11, for a variety of possible dilution with gas having the initial metallicity of the first generation. In Table\,\ref{table:1} we provide an estimate of the dilution required to obtain an abundance within the fiducial or the conservative error box. 
\begin{table}
\caption{Age of the Universe at the time of formation of the system progenitor of GN-z11
if M$_{AGB}$ is polluting the medium at age 430\,Myr}
\centering 
\begin{tabular}{c | c | c c | c c }
\hline
 Z & M$_{AGB}$ & \multicolumn{2}{c}{dilution \%} &  AGB age  & Age*  \\
               & M$_\odot$   & fiducial &  conservative      &  (Myr) & (Myr)            \\
\hline
\hline
0.001          & 7.5 & X  & 0--40   &  46.2  &   384   \\
             &     7.0 &    X  &  10--60   & 52.5   &    378       \\
             &      6.0 &     X &  60--80 & 70.8   &       360       \\
             &      5.0 &       X   & 30--60   & 102  &      330        \\
            &      4.0 &         X          & 10--30   & 168  &       262        \\
\hline
\hline
0.002     &     7.0 &    60--80  &  0--60   & 51.4   &    379      \\
             &      6.0 &    60--80 &  0--60 & 69.8   &       360       \\
             &      5.0 &    40--60   & 0--60   & 100 &      330        \\
            &      4.0 &          60         & 60   & 167  &       263        \\
\hline
\hline 
0.004    &      6.0 &    X         & 0--20 & 70.1   &       360      \\
             &      5.0 &    0--10   & 0--10   & 102  &      328        \\
            &      4.0 &     X         &  X   & 169  &       261       \\
\hline
\end{tabular}
\label{table:1}
\end{table}

\section{Can GN-z11 be a ``typical" second generation formation stage?}
\label{sec4}
When we consider the role of AGB ejecta and the dilution with gas having the composition of the first generation stars,  are we implicitly assuming to be dealing with the formation in a typical GC? The presence of a massive BH with a mass $\sim 10^6 M_{\odot}$\ is a clear indication that that is not the case and that the system investigated is instead likely to be the progenitor of a NSC. In the rest of this section we discuss other aspects that may differentiate the evolution of a typical GC from that of the GN-z11 massive star cluster.\\
{\bf 1) Are SN\,II ejecta of the first generation expelled} out of the system, so that the second generation may form without contamination from the supernova ejecta of the first generation? In the original AGB model by \citet{dercole2016}, it was proposed that the central bubble powered by the SN\,II breaks out of the disk of the host (dwarf) galaxy forming a funnel through which the ejecta are lost. But  if we apply the model to the physical conditions in GN-z11 central regions {\it (derived from the spectral energy distribution by models not considering the AGN component emission)}, the stalling radius\footnote{ Albeit a little inappropriately, bubbles whose radiative losses are negligible in front of the mechanical luminosity $L_{w}$ of the SN wind are called ``energy conserving";  in the opposite case the bubble shell is directly driven by the wind pressure and is called ``momentum conserving". For the energy conserving and momentum conserving bubbles the stalling radii are given, respectively, by $R_{st,e}=48\left ({L_{36}\over n} \right)^{0.5}$ pc and $R_{st,m}=5\left ({L_{36}\over n} \right)^{0.5}$ pc, where the mechanical luminosity of the wind is given in units of ${10}^{36}$ erg/s and $n$ is the numerical density of the ISM in cm$^{-3}$,  
%\citep[][and references therein]
see \citet{dercole1992} and references therein, but note that here the numerical coefficients are a bit lower because we do not consider the contribution of stars more massive than $25 M_{\odot}$. } (where the velocity expansion equals the sound speed of the surrounding ISM and the bubble stops its growth)
%Assuming, e.g., a GC with an initial stellar mass $M_{FG}=10^6 M_{\odot}$ we obtain, as a rule of thumb, $L_{36}=10^4$. Considering the density surrounding ISM in the range $10^2<n<10^4$ cm$^{-3}$ (Bunker et al. 2023), we obtain, respectively, $R_{st,e}=480$ pc and $R_{st,m}=50$ pc for $n=10^2$ cm$^{-3}$, and $R_{st,e}=48$ pc and $R_{st,m}=5$ pc for $n=10^4$ cm$^{-3}$. Excluding this last possibility (because the volume doesn't contain enough mass) we conclude that, in order to form a funnel, the thickness of the host galaxy disk must be in the range $500-50$ pc.
may become too small, and the ejecta remain in the cluster if the densities are larger than  $\sim 100$\,cm$^{-3}$ \citep[the lower range of the values given by][]{bunker2023}. The model is fully excluded if the densities were $n \simeq 10^5$cm$^{-3}$ \citep{senchyna2023}, because the stalling radius would be only 17\,pc, but also because the PS volume would contain more than 20$\times 10^6$\Msun, and such a value seems indeed too much for a second generation made up by AGB ejecta. 
It is important to keep in mind that the density values so far derived do not consider the AGN. If the massive cluster in GN-z11 is a NSC, the possible metallicity variations produced as a result of the retention of SN ejecta are not a problem.\\
{\bf 2) Is standard ``dilution" feasible?}
In spite of the presence of a central AGN emission, we maintain the idea that a first generation cluster is also present: it is needed both to produce the initial BH seeds and a favourable environment for their merging, and to obtain a cooling flow and star formation in a concentrated region. Observationally, most galactic GCs can still be considered mono-metallic, since the upper limit to the scatter of iron is less than $\sim$0.05\,dex   \citep{carretta2009c}.  Only recently a somewhat larger spread has been isolated in the first generation stars of some GCs \citep{legnardi2022}, but the strict homogeneity of the second generation stars (in spite of the variations in p-capture elements abundances) has been confirmed. This latter result points to a homogeneous mixing of the gas producing the second generation stars, probably due to its formation in a very compact region, such as the cooling flow, an indirect confirmation of the AGB model. \\
In the case of GN-z11 the chemical evolution and dynamics can be considerably different. In particular, we are at such a young  age that the Universe is recently emerging from the dark epoch, so the proto--galaxy or proto--cluster is probably still surrounded by dense neutral gas, which may take part in the accretion process. Further, AGB ejecta may be accreting from a larger region than the first generation cluster, and the ongoing event of star formation may produce also exploding and polluting core collapse supernovae. At the same time, the surrounding gas has probably a metallicity smaller than the pristine gas of the first generation stars, and its accretion may counteract the overall medium metallicity increase due to the recent supernovae \citep[see, e.g.][]{heintz2023}. If instead the gas composition is more metallic due to supernovae, it is possible that the first generation was less metallic than assumed in the simple dilution, putting back into play the models at Z=0.0002 we had dismissed.  Metallicity variations in large fractions of the stars are present in a few galactic GCs, like $\omega$\,Cen,  which in fact are often regarded as remnants of NSCs, as GN-z11 could be. In summary, the dilution with  gas having the ``same composition of the first generation stars", as shown in the Fig.\,\ref{fig:1}, is probably an oversimplified view, but, for what matters in the present context, since most of the ejecta compositions are considerably N-rich, any more complex dilution of the ejecta will be generally compatible with the GN-z11 spectrum, albeit possibly requiring a bit different initial composition to the first generation stars.\\
{\bf 3) How does accretion going on onto the central BH modify star formation?}
Accretion on the central BH puts another difference between the ``standard" AGB model and the present case. The AGN  injects energy into the surroundings through the process of accretion-driven feedback, enhancing the velocity dispersion of the gas in the cooling flow and/or driving the fast gas outflows actually seen in the spectrum \citep{maio2023bh106}. The problem is discussed in \citet{vesperini2010bh}. It is possible that accretion is intermittent, and that star formation is active mainly during the quiescent phases of the BH. On the other side, intermittency lengthens the time required to accumulate a  BH  mass of 10$^6$\Msun\ \citep{milosavljevic2009}.
One point to notice is that in this NSC environment, a favourable circumstance is that the AGB ejecta may be accreting from a larger region than the first generation cluster, as the high N/O is similar for several stellar masses, and a strict coevality is not necessary for the polluting AGB gas to show the N/O observed in GN-z11.

In conclusion, while we are far from a conclusive ``model" for the possible formation of a second generation population in GN-z11, the main point now is that we are witnessing the role of massive AGB winds in polluting a medium where a massive globular cluster is present, even if the final system emerging will be more complex than a standard GC.
\begin{figure}
\vskip -60pt  
\begin{minipage}{0.48\textwidth}
\resizebox{1.\hsize}{!}{\includegraphics{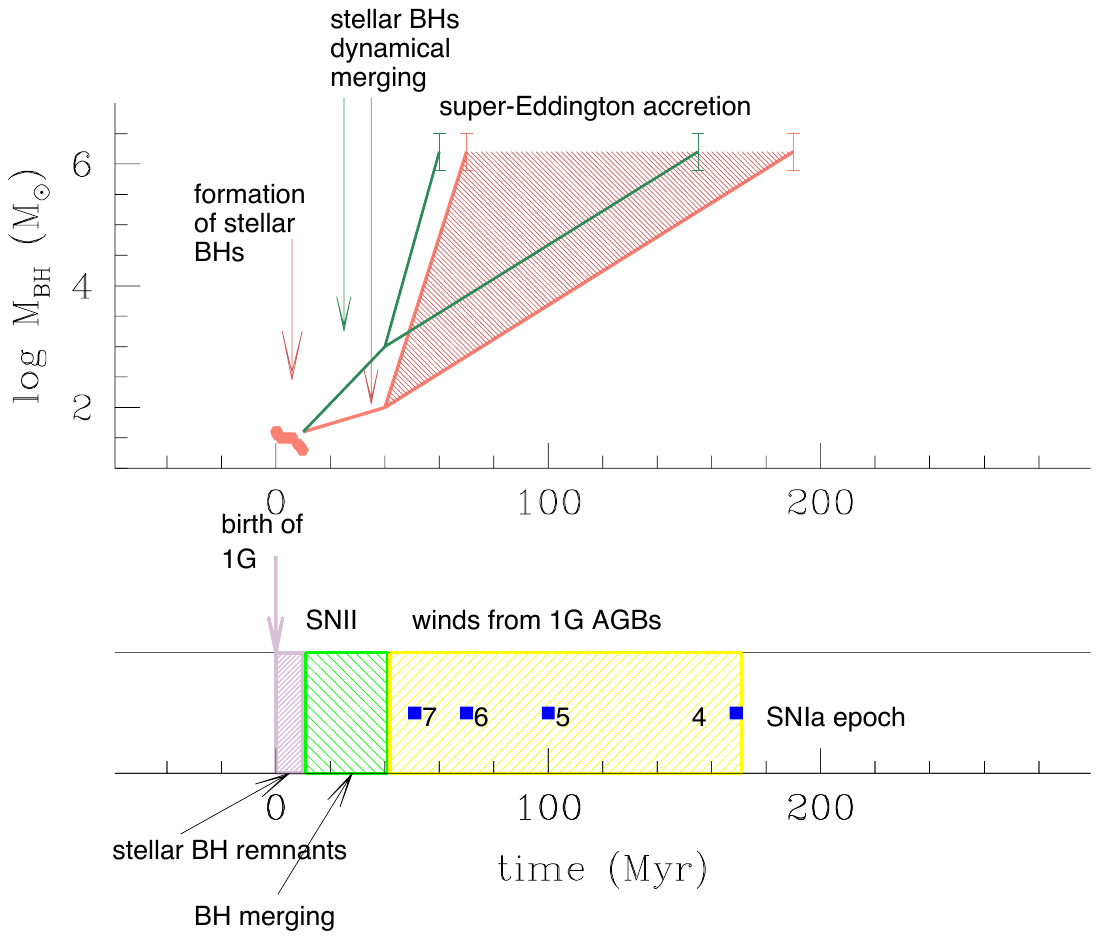}}	
\end{minipage}
\vskip-100pt
\caption{ A sketch of the players in the game. At the bottom we show the "ruler" of timescales in a standard Globular Cluster in the AGB model and at the top we show the possible growth of the central BH.
The time origin is set at birth of the first generation (1G). {\it GREY box}: formation of BHs of 40--20\Msun from the evolution of the most massive stars down to $\sim$25\Msun ($\sim$10\,Myr);   {\it GREEN box}: SN\,II epoch, scarce presence of gas;  stellar BHs merge and leave a seed BH of 100--1000\Msun (red and green lines in the upper panel) ($\sim$30\,Myr). {\it YELLOW box}: ``quiet" period of AGB evolution, the evolving masses are marked as blue squares. AGB ejecta and the re-accreting interstellar gas mix and produce the cooling flow giving origin to the formation of the second generation. Part the gas collecting into the central regions also feeds the BH ($\lesssim$130\,Myr). Afterwards, SN\,Ia begin exploding in the cluster. In the upper panel of the figure we show schematically how Super-Eddington accretion allows the BH to reach the mass of log(M/M$_\odot)=6.2 \pm 0.3$, starting from a BH of 100\Msun\ (red triangle) or 1000\Msun\ (green lines). Accretion occurs at the rates suggested from the AGN luminosity, following \cite{maio2023bh106}.}
\label{fig:2}       
\end{figure}
\begin{figure}
\vskip -60pt  
\begin{minipage}{0.48\textwidth}
\resizebox{1.\hsize}{!}{\includegraphics{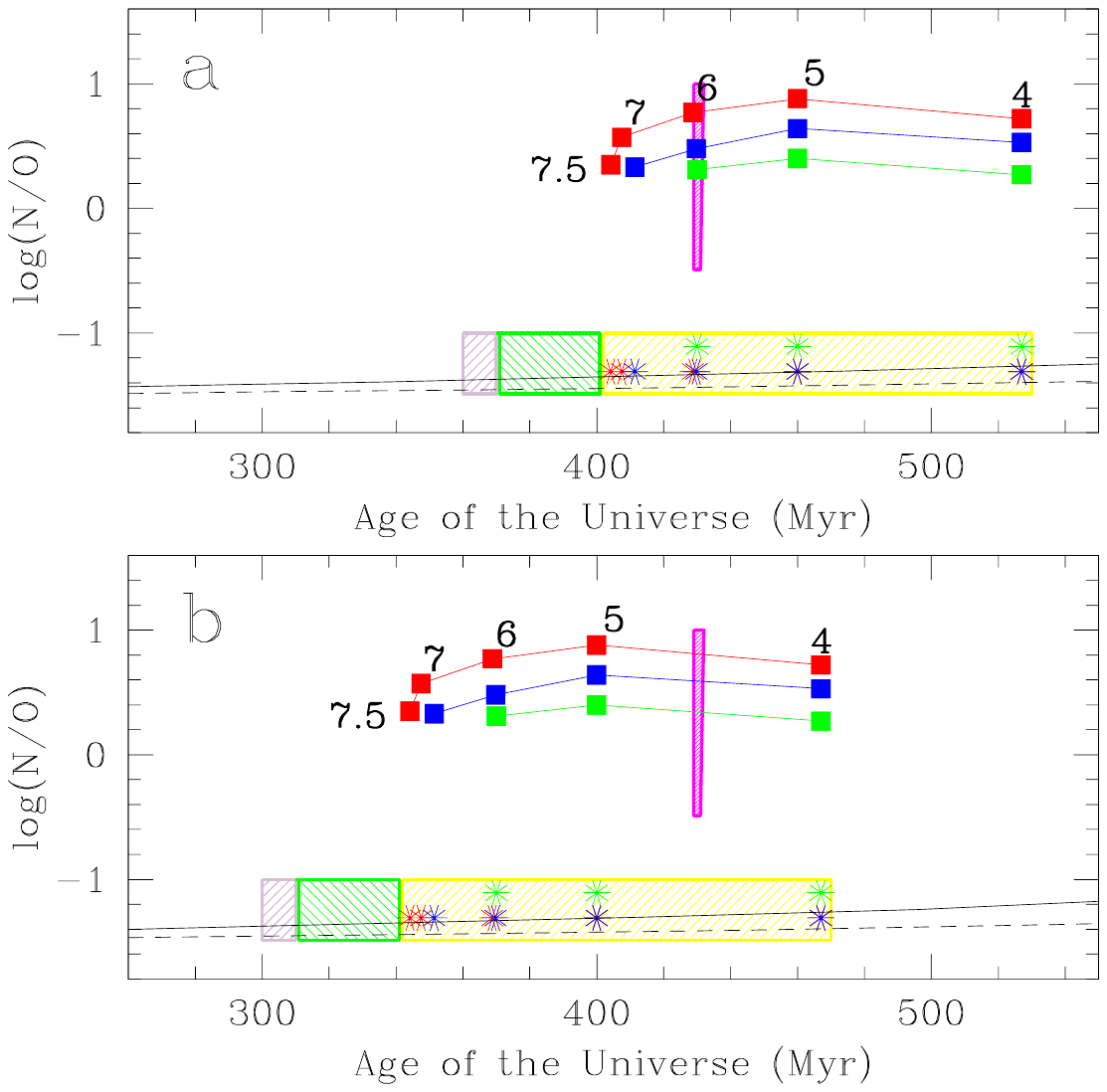}}	
\end{minipage}
\vskip-75pt
\caption{Two possible examples of the evolution in the proto--nuclear star cluster GN-z11, in the hypothesis that the AGB winds mixed with standard interstellar gas constitute the gas responsible for the high (N/O) abundances in the spectrum. Any of the intermediate mass AGB stars may pollute the intra cluster medium with the correct (N/O) (properly diluted with pristine gas). The pink rectangle is the fiducial error box by \cite{cameron2023} for GN-z11.  In Panel a we assume that the 6\Msun\ is the mass presently evolving, while in Panel b we assume that the 4.5\Msun\ is evolving, at the age of 430Myr. The formation of the 1G is shifted back by 70\,Myr (the evolution age of the 6\Msun) in Panel a and by 130\,Myr (the evolution age of the 4.5\Msun), respectively at $\sim 360$\,Myr  or $\sim$300\,Myr, as shown by the ``ruler" drawn at the bottom.  Asterisks denote the initial $\log$(N/O) in the models, and colored squares represent the abundance in the ejecta. If we were witnessing  the evolution of a larger (smaller) mass, the AGB abundances and the boxes would be all shifted to larger (smaller) age. The two lines at the bottom represent the time evolution of log(N/O) for the upper (full line) and lower (dashed) envelope of the chemical evolution models by \cite{vincenzo2016}.}
\label{fig:3}       
\end{figure}

\section{Examples of the timing to the present stage of GN-z11}
\label{sec5}
Contrary to previous discussions in the literature, we are claiming that the main point of advantage in identifying as massive AGB the polluters of the gas in GN-z11 is the relatively long timescale on which they act. In Fig.\,2 we sketch the ``ruler" of the interval of time between the formation of a first generation of stars and the end of the massive AGB phase.\\
During the first 10\,Myr, the most massive stars down to $\sim$25\Msun\  end their life into black hole remnants (possibly without exploding as SN) (grey box).
For all other models invoking the second generation star formation, the timeline of the event must be confined within the first 2--3\,Myr or at most within the 10\,Myr window indicated by the grey box.
%into this grey box, or is even squashed on the origin.  
After that, the explosions of SN\,II goes on for a long period ($\sim$30\,Myr), during which star formation is either totally suppressed because most of the gas is expelled out of the cluster, or it proceeds at a slower pace, while the dynamical interactions of the just formed remnant BHs in the central region lead to merging events beginning to increase the seed BHs. When the stars begin evolving into white dwarfs, and the SN\,II epoch ends, the interstellar gas and the N/O rich AGB ejecta begin falling back to the GC core, with the double effect of accreting gas on the central BH and of forming the second generation stars. Going on with the time, sequentially smaller masses evolve. Here we limit the time span to the evolution of the 4\Msun\ at $\sim$170\,Myr, so fully including the extreme period in which  s--Fe--anomalous clusters are formed \citep{dantona2016}. \\
Practically all the intermediate mass AGB stars (especially for Z=0.002) may pollute the intra cluster medium and provide the observed (N/O), but we can not know which mass is now contributing, at the age of 430\,Myr. The formation of the pristine GC will go back in the past by  50\,Myr (if now the $\sim$7\Msun\ is evolving) to 170\,Myr (if the 4\Msun\ is presently evolving) and the formation of the first generation is pushed back to $z \simeq$15--12 and ages of the Universe of only 260--380\,Myr  (column 6 in Table\,\ref{table:1}), fully into the pre--reionization epoch, leaving enough space for the first galaxies to have already formed \citep{robertson2022, harikane2022, calura2022}.  Positioning our ruler by imposing that a particular mass is evolving now in GN-z11 at the Universe age of 430\,Myr, we visualize the age of formation of the first generation and the timespan for accreting the central BH in GN-z11. Two examples are shown in the panels of Fig.\,\ref{fig:3}, in the hypothesis that the AGB ejecta of the 6\Msun\ (or 4.5\Msun) are in the accreting gas responsible for the high (N/O) abundances in the spectrum.  The 6\Msun\  (4.5\Msun) takes $\sim 70$\,Myr  ($\sim$130\,Myr) to evolve, so the first generation stars were born at $\sim$360\,Myr ($\sim$300\,Myr).  After 40\,Myr the accretion phase begins. Confronting the available times with the upper panel in Fig.\,\ref{fig:2} we see that in both cases the BH may reach the value of the BH in GN-z11 if subject to super-Eddington accretion.

\section{Conclusions}
\label{sec6}
We examine the possibility that the high N/O in the spectrum of GN-z11 is due to the presence of CNO cycled gas forming a second generation in a massive Globular Cluster. The conclusions are:
\begin{enumerate}
\item GN-z11 morphology,  points either to a very concentrated star formation event and/or to AGN emission in matter enriched by N/O rich ejecta;

\item we have shown that the ejecta of massive AGB of a first generation Globular Cluster, mixed with the interstellar matter, provide values of N/O consistent with the abundances in the spectrum of GN-z11. The scenario is probably more complex than in a standard GC second generation formation:  as  the emission of the central point of GN-z11 is probably due to an AGN type emission, from a BH mass of $\sim$10$^6$\Msun\ \citep{maio2023bh106}, the cluster resembles more a Nuclear Star Cluster than a standard GC;

\item the models of formation of the second generation proposed till now are active for a time span of $\sim$1\,Myr to $\sim$10\,Myr. The time span of action of the AGB model is much more long lived, 
and may extend up to 130 Myr;

\item the presence of a first generation cluster which has formed many stellar BH as remnants in the first Myr's of life is critical to produce a central seed BH with mass of $\sim$100--1000\Msun,  through mergers in the system dense central environment;

\item the long duration of the massive AGB phase of pollution and of the associated cooling flow is 
a favourable environment for gas accretion on a seed BH and to increase its mass to the high value proposed for the GN-z11 BH, at an age of the Universe as young as 430\,Myr, without having to resort to the hypothesis of a direct collapse leading to the rapid formation of such a massive BH.
\end{enumerate}
In summary, we proposed a scenario to explain a few fundamental properties of the GN- z11 system observed at z=10.6 by JWST. Our study shows that the observed properties of this system are consistent with a model combining the formation of second generation stars with large N/O out of pristine gas and AGB ejecta and the growth of a massive BH from gas accretion on a central seed BH. Star formation and the growth of super massive BHs are key ingredients in galaxy formation  and 
in future studies, it will be fundamental to understand the role of each component in defining the spectrum of GN-z11 and other similar systems at high redshift.

\label{end}

\begin{acknowledgements} 
EV acknowledges support from NSF grant AST-2009193.
FC acknowledges funds from the European Union -- NextGenerationEU within the PRIN 2022 project n.20229YBSAN - Globular clusters in cosmological simulations and in lensed fields: from their birth to the present epoch. PV acknowledges support from the National Institute for Astrophysics (INAF) within the theory grant ``Understanding mass loss and dust production from evolved stars'' (ObFu 1.05.12.06.07, PI. P. Ventura) and the PRIN 2019 grant `Building up the halo: chemodynamical tagging in the age of large surveys?'' (ObFu 1.05.01.85.14, PI. S. Lucatello). AFM, APM,  FDA and FD acknowledge support from the INAF Research GTO Grant 2022 RSN2-1.05.12.05.10 ``Understanding the formation of globular clusters with their multiple stellar generations'' (P.I. A. Marino) of the ``Bando INAF per il Finanziamento della Ricerca Fondamentale 2022". AFM and APM also acknowledge support  by MIUR under PRIN program 2017Z2HSMF (PI: Bedin).

\end{acknowledgements}

% WARNING
%-------------------------------------------------------------------
% Please note that we have included the references to the file aa.dem in
% order to compile it, but we ask you to:
%
% - use BibTeX with the regular commands:
%   \bibliographystyle{aa} % style aa.bst
%   \bibliography{Yourfile} % your references Yourfile.bib
%
% - join the .bib files when you upload your source files
%-------------------------------------------------------------------
\bibliographystyle{aa}
\bibliography{gnz11}

\end{document}